# Modélisation et extraction de données pour un entrepôt objet

**Franck Ravat, Olivier Teste, Gilles Zurfluh**

Université Paul Sabatier (Toulouse III),
IRIT (Institut de Recherche en Informatique de Toulouse), équipe SIG,
118, Route de Narbonne - 31062 Toulouse cedex 04, France

*tel* : (0)5 61 55 63 22,
*mel* : {ravat, teste, zurfluh}@irit.fr

**Résumé :** Cet article traite de la modélisation orientée objet pour la conception d'un entrepôt de données complexes et historisées (conservation des évolutions). Un des aspects majeurs de cette modélisation est l'extension du concept de classe par celui de classe entrepôt, défini au travers d'un filtre temporel et d'un filtre d'archives ainsi que d'une fonction de construction. Les filtres gèrent l'évolution des données sous une forme pertinente (détaillée ou archivée) pour l'aide à la décision. La fonction de construction définit la structure et les données des classes entrepôt à partir d'un processus d'extraction appliqué sur une source globale intégrée. L'intérêt de notre processus d'extraction est qu'il combine les structures et le comportement des données. L'extraction du comportement des classes s'effectue au travers du concept de matrice d'usage.

## 1 Introduction

Les entrepôts de données ("*data warehouses*") permettent de stocker l'ensemble des informations nécessaires à la prise de décision afin d'améliorer les choix stratégiques effectués par les décideurs dans les entreprises. Les entrepôts sont alimentés par des extractions de données appliquées à des sources d'informations comme les bases de production [WIDO95] [INMO96] [JARK99].

Nos travaux de recherche, en continuité de ceux effectués dans notre équipe en conception de systèmes d'information en milieu hospitalier [LAPU97], visent à élaborer un système d'aide à la décision, basé sur l'approche des entrepôts de données, dans le milieu médical[1]. L'objectif de ce système est d'améliorer l'analyse, le suivi et le contrôle des dépenses de santé, de l'activité des médecins et des patients. Ces travaux se situent également dans le cadre du projet REANIMATIC[2].

Nos premiers travaux nous ont permis de définir l'architecture fonctionnelle d'un système d'aide à la décision [BRET99], décomposée en trois niveaux (intégration, construction, structuration), distinguant les problématiques de recherche.

- L'**intégration** se propose de résoudre les problèmes d'hétérogénéité (modèles, formats et sémantiques des données, systèmes,…) des différentes sources de données (pouvant être des bases relationnelles, des bases orientées objet, des fichiers textes…) en intégrant celles-ci dans une source globale. La source globale est décrite au moyen du modèle de données orientées objet standard de l'ODMG [CATT95]. Le choix du paradigme objet se justifie car il s'avère parfaitement adapté pour l'intégration de sources hétérogènes [BUKH93] couramment utilisées dans le milieu médical [PEDE98]. Cette source globale est virtuelle, c'est à dire que les données utilisées pour la décision restent stockées dans les sources de données et sont extraites uniquement au moment des mises à jour de l'entrepôt. L'intégration s'appuie sur des techniques de bases de données fédérées [SAMO98] et réparties [RAVA95].

---

[2] *Le projet* ***REANIMATIC****, en collaboration avec l'association de médecins OUTCOME-REA, vise à concevoir et à développer un entrepôt de données médicales et évolutives (collectées à partir des bases opérationnelles des services de réanimation) afin d'améliorer la qualité des soins et le devenir des patients dans les services de réanimation des hôpitaux français.*

- La **construction** consiste à extraire les données pertinentes pour la prise de décision, puis à les recopier dans l'entrepôt de données, tout en conservant, le cas échéant, les changements d'états des données. Par conséquent, l'entrepôt de données constitue une collection centralisée, de données matérialisées et historiques (conservation des évolutions), disponibles pour les applications décisionnelles. Le modèle de l'entrepôt décrivant ses données doit supporter des structures complexes [PEDE98] et supporter l'évolution de ses données au cours du temps [INMO96] [PEDE98] [YANG00].
- La **structuration** réorganise l'information décisionnelle dans des magasins de données afin de supporter efficacement les processus d'interrogation et d'analyse, tels que les applications OLAP ("*On-Line Analytical Processing*" [CODD93]) et la fouille de données ("*Data Mining*" [FAYY96]) ; les industriels proposent de nombreux outils permettant une telle activité (*Oracle Express, Business Object, SAS, Impromptu,...*). Pour ce faire, les données importées dans les magasins sont souvent organisées de manière multidimensionnelle [AGRA97] [BRET00].

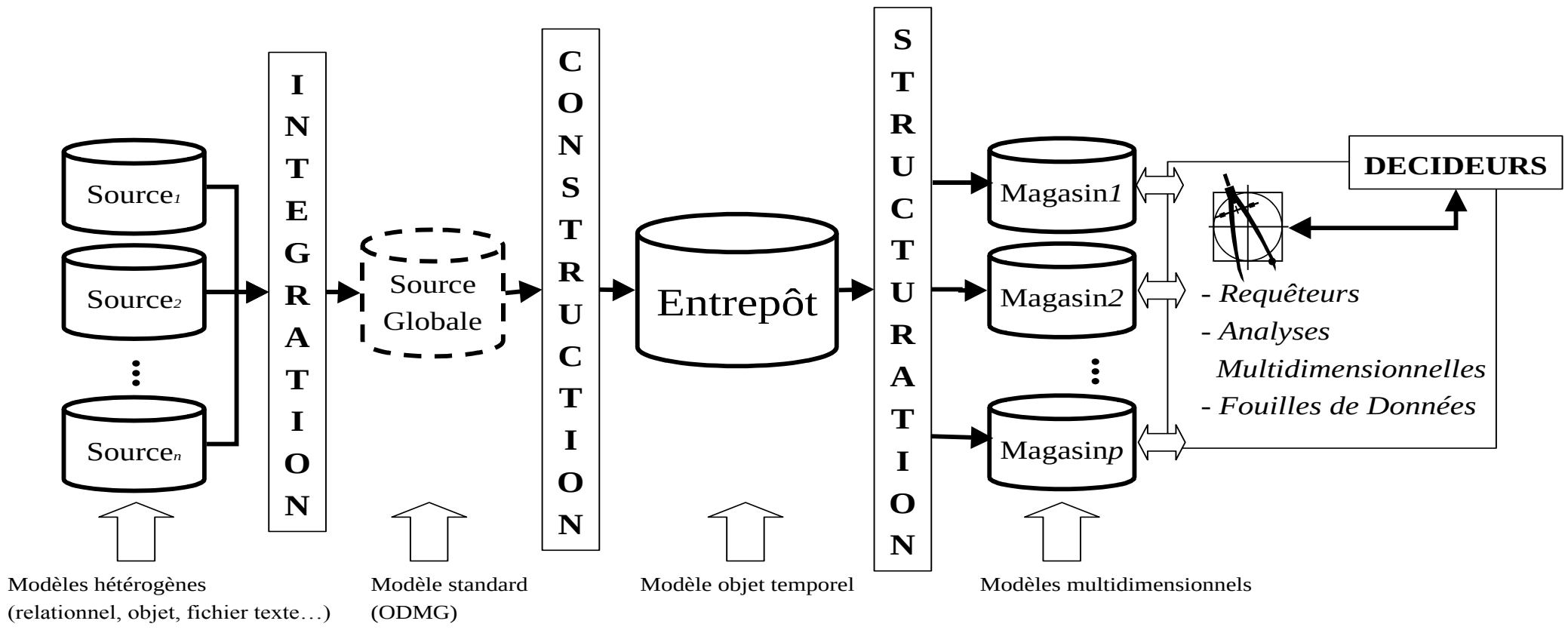

Figure 1 : *Architecture fonctionnelle d'un système d'aide à la décision.*

Il est important de remarquer que notre entrepôt de données n'est pas organisé de manière multidimensionnelle puisqu'il ne supporte pas directement les processus décisionnels (OLAP). Cette activité est réservée aux magasins de données qui améliorent les performances d'interrogation sans se soucier des redondances d'information ; chaque magasin stocke une partie de l'information disponible dans l'entrepôt afin de répondre à un objectif décisionnel précis ou à un groupe d'utilisateurs ayant les mêmes besoins.

Cet article se focalise sur le second niveau de notre système et aborde plus particulièrement le problème de la **conception de l'entrepôt**. Cependant, en annexe de cet article (section 8), nous proposons un exemple complet constitué de trois sources de données, d'une source globale, d'un entrepôt de données et un magasin de données. La source globale servira de base pour les exemples illustrant l'article.

L'entrepôt de données est élaboré à partir d'une source globale intégrant plusieurs bases de données de production ; il constitue une collection d'informations pertinentes pour aider la prise de décision dans des organisations comme les entreprises. Par conséquent, cela nécessite une modélisation spécifique répondant aux besoins décisionnels [INMO96] [CHAU97] ; en particulier, l'entrepôt de données doit

- Supporter des données à structures complexes qui sont couramment employées dans notre contexte d'application (le milieu médical [PEDE98]), tout en limitant la redondance d'information et en assurant l'intégrité des données afin de garantir la pérennité dans le temps de l'information décisionnelle,
- Proposer des mécanismes d'extraction permettant de recopier et de reformater les données issues de la source ainsi que leurs traitements associés afin de générer l'entrepôt,

- Supporter des données temporelles puisque les applications décisionnelles sont basées sur l'information courante, mais également sur l'histoire des données (évolutions successives), tandis que les sources de données ne sont pas adaptées à cette activité [INMO96] [CHAU97] [YANG98] [YANG00]. La conservation de l'historique des données doit être réalisée de manière pertinente (détaillée ou résumée) selon les besoins des utilisateurs.

Dans un premier temps, nous proposons **un modèle de données pour entrepôts de données complexes et temporelles**, puis dans un second temps nous définissons des **mécanismes d'extraction des données** traitant de l'aspect statique des données (**les données et leurs structures**) et de l'aspect dynamique (**le comportement des données**).

Un état de l'art sur le domaine des entrepôts est présenté dans la section suivante. La section 3 présente notre modèle de données orientées objet dédié aux entrepôts de données. La section 4 traite de l'extraction des données et de leurs structures tandis que la section 5 étudie l'extraction automatique du comportement des données.

## 2 Recherches sur les entrepôts de données

Issus de l'industrie, les entrepôts de données sont aujourd'hui un thème de recherche à part entière. L'approche multidimensionnelle [AGRA97] [GYSS97] [LEHN98] [PEDE99] [BRET00] et l'approche des vues matérialisées [WIDO95] [GUPT95] [YANG97] [YANG 98] [THEO99] [YANG00], déterminant essentiellement les performances de l'entrepôt, font l'objet de nombreux travaux, tandis que les aspects plus conceptuels restent peu étudiés [GATZ99].

La plupart des travaux actuels se contentent d'utiliser des vues relationnelles matérialisées [GUPT95] pour élaborer les entrepôts. Or, de nombreuses applications telles que les applications médicales, nécessitent des **modèles de données plus réalistes et sémantiquement plus riches** que le modèle relationnel [PEDE98] [PEDE99] [GOPA99]. De nouvelles propositions vont dans ce sens, notamment, [GOPA99] qui développe un mécanisme de vues matérialisées basées sur le modèle objet-relationnel ainsi que [PEDE99] qui définit un modèle orienté objet multidimensionnel pour données complexes dans milieu médical. Tandis que [GOPA99] n'aborde pas le problème essentiel de la gestion du temps, [PEDE99] propose un modèle temporel visant à ajouter un temps de validité aux données (en stockant l'information temporelle dans une table de dimension).

D'autres travaux [YANG98] [YANG00], dans le cadre du projet WHIPS ("*WareHouse Information Project at Stanford*"), intègrent le temps en proposant des algorithmes basés sur la technique des vues matérialisées définies au travers d'un langage équivalent à TSQL2. Ces travaux intègrent l'information temporelle sous la forme d'un attribut temporel ajouté aux tables. Les propositions actuelles concernant la gestion du temps dans les entrepôts n'offrent pas de mécanismes flexibles permettant l'**archivage des données temporelles** ; le plus souvent, les données anciennes sont simplement supprimées de l'entrepôt [YANG00].

Enfin, les travaux actuels traitent principalement de l'extraction des données, mais l'étude de l'extraction des comportements des données n'est pas abordée. En particulier, les dernières propositions développant des modèles intégrant des structures de données à objet [PEDE99] ou objet-relationnel [GOPA99] n'abordent pas l'extraction du comportement. En outre, [GOLF99] propose une méthode permettant de transformer un schéma entité/association en un schéma en étoile sans intégrer le comportement des entités. Dernièrement, dans [CUI00] les auteurs abordent les problèmes liés à la correspondance entre les données stockées dans l'entrepôt et les données source dont elles sont issues (*the tracing lineage problem*). Ces travaux se placent dans le contexte d'un entrepôt de données défini par des vues relationnelles matérialisées ; ils ne traitent pas de l'**extraction du comportement des données**.

## 3 Modèle de données de l'entrepôt

Dans cette section, nous définissons un modèle de données pour les entrepôts, basé sur le paradigme objet. Notre modèle subit l'influence du modèle objet standard de l'ODMG [CATT95] qui est étendu pour prendre en compte les caractéristiques des entrepôts de données. Notamment, notre modèle intègre la dimension temporelle d'une manière flexible en permettant l'archivage des données temporelles.
La section 3.1 introduit le concept d'objet entrepôt. La section 3.2 présente le concept de classe entrepôt. Enfin, la section 3.3 définit le concept d'environnement.

## 3.1 Objet entrepôt

Chaque information extraite (objet, partie ou groupe d'objets source) est représentée dans l'entrepôt par un **objet entrepôt** qui conserve ses évolutions de valeur au cours du temps (tandis que la source de données ne contient que l'état courant [CHAU97], ou bien, ne conserve qu'une partie récente des évolutions, insuffisante pour la prise de décision [YANG00]). Dans un entrepôt, l'administrateur peut décider de conserver :

- l'image de l'information extraite c'est à dire l'**état courant**, ainsi que
- les états successifs que prend l'information extraite dans le temps, c'est à dire ses **états passés**,
- uniquement un résumé des états passés successifs, c'est à dire l'agrégation de certains états passés, appelée **état archivé**. Les états passés ainsi résumés sont supprimés de l'entrepôt afin de limiter l'accroissement du volume des données.

Un **objet entrepôt** est donc défini par le quadruplet ($oid$, $S_0$, $EP$, $EA$) où $oid$ est l'identifiant interne, $S_0$ est l'état courant, $EP = \{S_{p1}, S_{p2},\ldots, S_{pn}\}$ est un ensemble fini contenant les états passés et $AP = \{S_{a1}, S_{a2},\ldots, S_{am}\}$ est un ensemble fini contenant les états archivés.
Un **état** $S_i$ d'un objet entrepôt est défini par le couple ($v_i$, $h_i$) où $v_i$ est la valeur de l'objet pour les instants de $h_i$ et $h_i = <[td^1, tf^1[;\ldots;[td^h, tf^h[>$ est le domaine temporel (ensemble ordonné d'intervalles disjoints deux à deux) définissant les instants durant lesquels la valeur de l'état $S_i$ est courante.
La modélisation des domaines temporels s'effectue au travers d'un modèle temporel, linéaire, discret qui définit le temps par le biais d'unités temporelles ; l'espace continu du temps, représenté par une droite de réels, elle-même décomposée en une suite d'intervalles consécutifs disjoints [FAUV99]. Chaque partition correspond à une unité temporelle caractérisée par la taille des intervalles décomposant la droite du temps. Notre modèle gère un ensemble d'unités temporelles nommées (*année*, *semestre*, *trimestre*, *mois*, *semaine*, *jour*, *jour_semaine*, *heure*, *minute*, *seconde*), muni d'une relation d'ordre partiel *est-plus-fine* permettant de comparer les unités. Nous définissons plusieurs types temporels de base : l'instant, l'intervalle ainsi que le **domaine temporel**. Ce dernier est un ensemble ordonné d'intervalles disjoints deux à deux et non contigus, noté $h_i = <[td^1, tf^1[; [td^2, tf^2[;\ldots;[td^h, tf^h[>$ où chaque intervalle est non vide ($\forall k \in [1..h]$, $td^k < tf^k$) et possède une même unité temporelle ($\forall k \in [1..h], \forall j \in [1..h]$, unit($[td^k,tf^k[$)=unit($[td^j,tf^j[$ tel que la fonction *unit*(*Int*) retourne l'unité temporelle de *Int*).

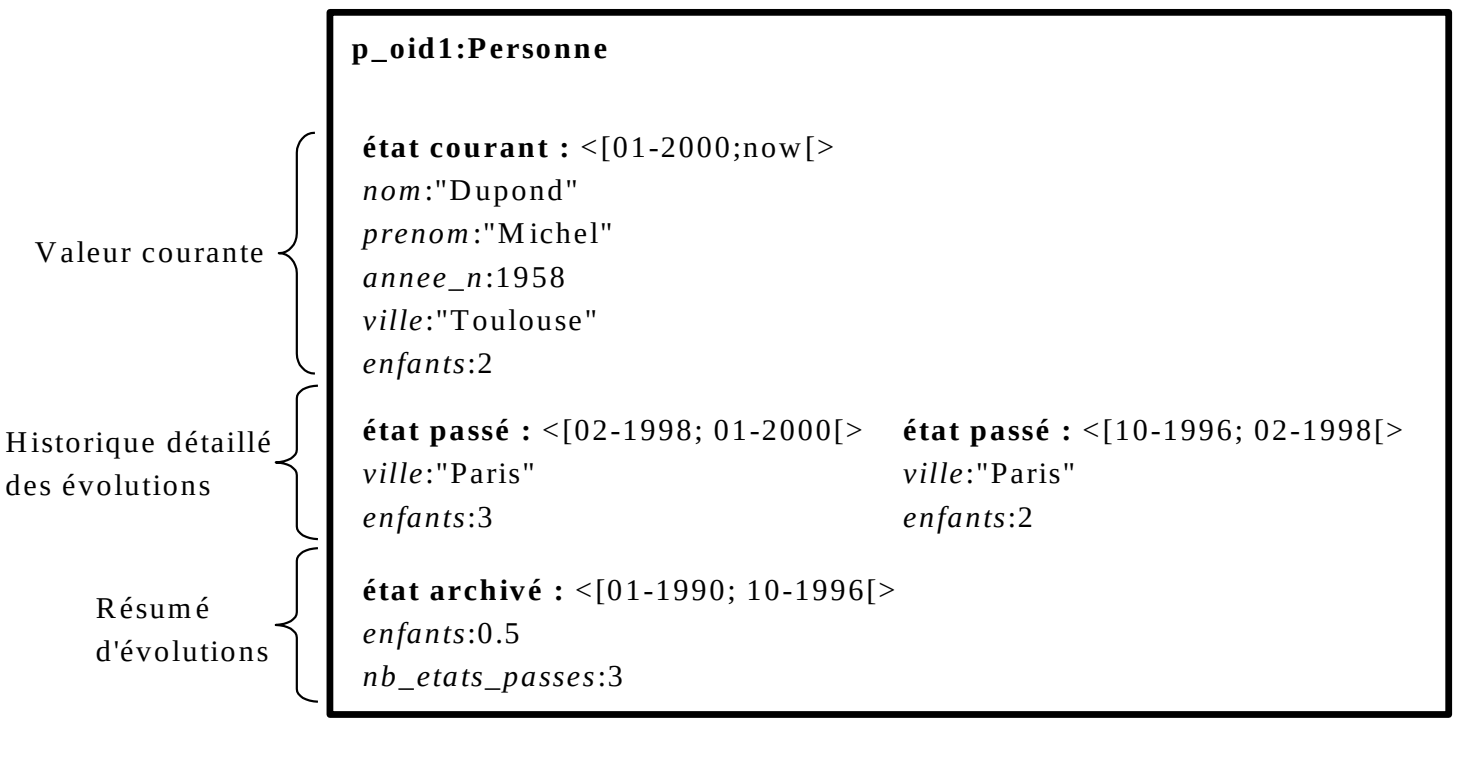

Figure 2 : *Exemple d'objet entrepôt.*

<u>*Exemple*</u> : Considérons un objet entrepôt représentant une personne décrite par un nom, un prénom, une année de naissance, la ville où elle réside et le nombre d'enfants à charge. La Figure 2 décrit un objet entrepôt identifié par "*p_oid1*" comprenant un état courant (qui représente la dernière valeur extraite pour chaque attribut), deux états passés (qui représentent le détail des évolutions) et un état archivé (qui résume trois états passés). Nous remarquons que les évolutions de toutes les propriétés ne sont pas nécessairement conservées ; ainsi sur la figure, les états passés conservent les évolutions des attributs "*ville*" et "*nb_enfants*" tandis que l'état archivé ne résume que les évolutions de l'attribut "*nb_enfants*".

## 3.2 Classe entrepôt

### 3.2.1 Définition

Les objets entrepôt qui ont la même structure et le même comportement, sont regroupés dans une classe. Pour prendre en compte les caractéristiques des objets entrepôt, nous définissons le concept de **classe entrepôt** *c* par un n-uplet constitué

- du nom de la classe $Nom^c$,
- d'un type $Type^c$ définissant la structure $Structure^c$ et le comportement $Comportement^c$ des objets entrepôt de *c* (à chaque classe entrepôt correspond un type),
- d'un ensemble fini de super classes $Super^c$ ($c_i$ est une super classe de *c*, notée $c \preccurlyeq c_i$ si et seulement si, $Type^c \supseteq Type^{ci}$ et $Extension^c \subseteq Extension^{ci}$),
- d'une extension $Extension^c = \{o_1, o_2, \ldots, o_x\}$,
- d'une **fonction de construction** $Mapping^c$ qui permet de spécifier le processus d'extraction et de transformation mis en jeu pour créer la structure et le peuplement de la classe *c* à partir de la source globale,
- d'un **filtre temporel** $Tempo^c$ définissant l'ensemble des propriétés temporelles de *c* (une propriété est temporelle lorsque ses évolutions sont conservées par des états passés). Le filtre temporel caractérise la structure des états passés des objets de la classe ($\forall prop \in Structure^c$, si $prop \in Tempo^c$ alors $\forall S_i \in Passe(c), prop \in Structure^{Si}$ où la fonction *Passe(c)* retourne l'ensemble des états passés de la classe *c* et $Structure^{Si}$ désigne la structure d'un état $S_i$).
- d'un **filtre d'archives** $Archi^c$ définissant l'ensemble des propriétés archivées de *c* (une propriété est archivée lorsque ses évolutions passées sont résumées dans des états archivés). Le filtre d'archives caractérise la structure des états archivés des objets de la classe ($\forall prop \in Structure^c$, si $prop \in Archi^c$ alors $\forall S_j \in Archive(c), prop \in Structure^{Sj}$ où la fonction *Archive(c)* retourne l'ensemble des états archivés de la classe *c*).

La fonction de construction fait l'objet d'une étude approfondie à la section 4.

<u>Exemple</u> : Reprenons l'exemple de la section 3.1. Le type de la classe "*Personne*" est défini par l'expression suivante :

```
interface Personne {
        attribute String nom;
        attribute String prenom;
        attribute Short annee_n;
        attribute String ville;
        attribute Short nb_enfants;
        }
with temporal filter {ville, nb_enfants},
     archive filter {avg(nb_enfants)}
```

Cette définition comprend un filtre temporel $Tempo^{Personne}$ et un filtre d'archives $Archi^{Personne}$ spécifiant respectivement les propriétés temporelles (définissant la structure des états passés) et les propriétés archivées (définissant la structure des états archivés).

### 3.2.2 Mécanisme d'archivage

Les propriétés archivées sont associées à une fonction d'agrégation qui indique comment sont résumées les évolutions détaillées de la propriété temporelle correspondante. Notre modèle supporte plusieurs catégories de fonctions d'agrégation :

- Les fonctions d'agrégation classiques (*avg, sum, count, max, min*). Ces fonctions résument les états passés sélectionnés pour l'archivage dans un seul état archivé.

- Les fonctions d'agrégation temporelles (*avg_t, sum_t, count_t, max_t, min_t*). Ces fonctions résument les états passés sélectionnés pour l'archivage avec plusieurs états archivés. Les états passés sélectionnés sont regroupés par grain de temps à une unité temporelle supérieure.

*Exemple* : La Figure 3 illustre la différence entre l'agrégation classique et l'agrégation temporelle. On suppose que l'unité temporelle des états passés et de l'état courant est l'unité *mois* et que l'archivage est appliqué à tous les états temporels non contenus dans l'année 2000.

- La fonction d'agrégation classique génère un seul état archivé qui regroupe l'ensemble des états temporels qualifiés pour l'archivage.
- La fonction d'agrégation temporelle génère plusieurs états archivés regroupant par année chacun les états temporels qualifiés pour l'archivage.

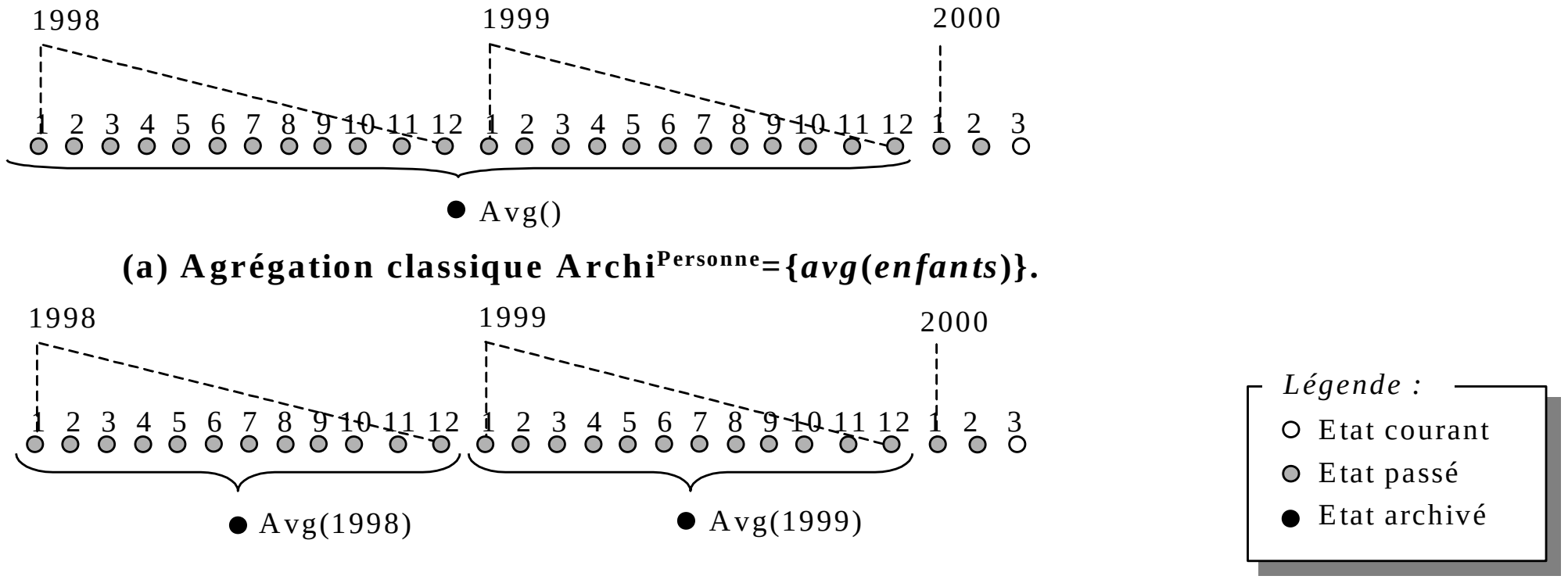

**(a) Agrégation classique Archi**$^{Personne}$**={*avg*(*enfants*)}.**

**(b) Agrégation temporelle Archi**$^{Personne}$ **={*avg_t*(*enfants, année*)}.**

Figure 3 : *Principes de l'archivage.*

L'avantage de notre modélisation est de permettre à l'administrateur de définir des classes entrepôt dont les objets conservent leurs évolutions sous une forme adéquate, c'est à dire en conservant soit le détail des évolutions (états passés), soit une agrégation de certaines évolutions dont le détail n'est plus utile (états archivés). Ce mécanisme réduit le volume des données temporelles dans l'entrepôt au strict nécessaire.

### 3.3 Environnements, Schéma de l'entrepôt, Configurations

Cette dualité (passé/archive) dans la conservation de l'évolution des données induit une nouvelle problématique relative au comportement temporel des classes. Il est indispensable de fournir des mécanismes permettant :

- de définir des critères pour caractériser les états passés à archiver,
- de garantir l'intégrité des relations sémantiques (associations, compositions) temporelles (dont les évolutions sont conservées). En effet, conserver l'évolution d'une relation exige de conserver les états passés impliqués dans la relation.

Pour cela, nous définissons le concept d'environnement constitué d'un ensemble de classes entrepôt ayant un comportement temporel homogène (même critère d'archivage, même périodicité de rafraîchissement…). Un **environnement** $Env_i$ est défini par un triplet constitué d'un nom $Nom^{Envi}$, d'un ensemble fini de classes de l'entrepôt $C^{Envi} = \{c^{Envi}_1, c^{Envi}_2,\ldots, c^{Envi}_{ni}\}$ et d'un ensemble de règles de configuration $Config^{Envi}$ visant à définir le comportement temporel de l'environnement.

Un environnement constitue donc une partie temporellement homogène dans l'entrepôt, ayant ses propres configurations locales $Config^{Envi}$. Nous avons étudié en détail dans [RAVA99] les environnements et leurs configurations. Toutefois, remarquons que ce concept d'environnement aide l'administrateur à définir différentes parties temporelles dans l'entrepôt, munies de leurs propres règles de configuration. Ceci permet de concevoir un entrepôt flexible qui s'adapte aux différentes exigences des décideurs.

Un entrepôt se caractérise par son **schéma** $S^{ED}$ défini par un nom $Nom^{ED}$, l'ensemble fini des classes de l'entrepôt $C^{ED} = \{c_1, c_2,\dots, c_n\}$, l'ensemble fini des environnements $Env^{ED} = \{Env_1, Env_2,\dots, Env_{ne}\}$ et un ensemble de règles de configuration $Config^{ED}$, visant à définir les différents paramètres de configuration globale de l'entrepôt (période de rafraîchissement,…).

# 4 Extraction des données

Cette section traite de la problématique concernant l'extraction des données pour construire des classes entrepôt. Le mécanisme d'extraction doit prendre en compte plusieurs exigences :

- dériver les structures de la source pour définir celles des classes entrepôt,
- peupler les classes entrepôt en recopiant uniquement l'information utile aux décideurs,
- transformer certaines structures de données extraites pour redéfinir des structures pertinentes pour l'entrepôt,
- organiser la hiérarchie d'héritage des classes entrepôt afin de structurer efficacement le schéma des classes de l'entrepôt.

Pour cela, chaque classe entrepôt est partiellement définie par une fonction de construction $Mapping^c$ appliquée sur la source de données (comme nous l'avons introduit à la section 3.2). Cette fonction est une composition de fonctions de base. Nous proposons une taxinomie des fonctions de base supportées par notre modèle d'entrepôt de données répondant aux différents problèmes posés :

- Les **fonctions de structuration** (FS) induisent la structure (attributs et relations) des classes entrepôt,
- Les **fonctions de peuplement** (FP) induisent les objets source à partir desquels l'extension des classes entrepôt est calculée,
- Les **fonctions ensemblistes** (FE) correspondent aux opérations ensemblistes classiques de l'algèbre objet de [SHAW90] en offrant des mécanismes puissants pour combiner et transformer les classes afin de constituer des classes entrepôt adaptées aux besoins des décideurs,
- Les **fonctions de hiérarchisation** (FH) organisent la hiérarchie d'héritage dans l'entrepôt en créant des super classes et des sous classes.

Nous posons $\forall c_i \in C^{ED}$, $Mapping^{ci} = f^{ci}{}_1 o f^{ci}{}_2 o \dots o f^{ci}{}_m$ avec $\forall j \in [1,m]$, $f^{ci}{}_j \in FS \vee f^{ci}{}_j \in FE \vee f^{ci}{}_j \in FP \vee f^{ci}{}_j \in FH$.

## 4.1 Fonctions de structuration

Les fonctions de structuration regroupent la fonction de projection ($\pi$), la fonction de masquage ($\mu$) et la fonction d'accroissement ($\alpha$) ; FS=$\{\pi, \mu, \alpha\}$. Leur rôle est de définir les propriétés constituant la structure des classes entrepôt.

La fonction de **projection** $\pi_{\{p1, p2,\dots,pp\}}(cs)$ dérive les propriétés de la classe source spécifiées dans l'ensemble $\{p_1, p_2, \dots, p_p\}$ et l'extension $Extension^c$ de la classe entrepôt $c$ est calculée à partir de tous les objets de la classe source *cs*. Inversement, la fonction de **masquage** $\mu_{\{p1, p2,\dots,pq\}}$ (cs) dérive les propriétés de la classe source non spécifiées dans l'ensemble $\{p_1, p_2, \dots, p_q\}$.

La fonction d'**accroissement** $\alpha_{\{p1:f1,\ p2:f2,\dots,\ pr:fr\}}(cs)$ permet de créer de nouvelles propriétés $\{p_1, p_2, \dots, p_r\}$, qui sont soit des attributs calculés, soit des propriétés spécifiques.

- Un attribut calculé représente une information, issue de la source de données, définie au travers d'une fonction appliquée à la source $f_i \in \{count, sum, avg, max, min\}$ ou d'une méthode source $f_i \in Comportement^{cs}$. Il représente explicitement une information contenue implicitement dans la source de données.
- Une propriété spécifique est définie par un type simple $f_i \in$ {*String*, *Short*, *Unsigned Short*, *Float*, *Double*, *Boolean*…} ou complexe $f_i \in$ {*Set(Type)*, *List(Type)*, *Struct(Type)*} connu dans l'entrepôt. Les propriétés spécifiques ne sont pas issues de la source ; elles sont ajoutées par l'administrateur dans l'entrepôt afin de donner la possibilité aux utilisateurs de compléter l'information de l'entrepôt en fonction de leur propre connaissance.

*Exemple* : Considérons la classe source "*PRATICIEN*" à partir de laquelle l'administrateur souhaite créer une classe entrepôt. Pour cela, il élabore une fonction de construction (utilisant la projection des propriétés utiles) comme suit.

```
π[o.nom,o.prenom,o.annee_n,o.diplome,o.adresse.ville] (o PRATICIEN)
```

La classe entrepôt générée est constituée des attributs dérivés "*nom*", "*prenom*", "*diplome*", "*annee_n*" et "*ville*".

L'administrateur peut également augmenter la classe entrepôt en utilisant une fonction d'accroissement imbriquée.

```
π[o.nom,o.prenom,o.annee_n,o.diplome,o.adresse.ville]
                                          (o α[nb_enfants:count(p.enfants)](p PRATICIEN)
```

Dans l'exemple, un attribut calculé "*nb_enfants*" augmente la classe ; il est obtenu en calculant le nombre d'enfants à charge pour chaque praticien.

## *4.2 Fonctions de peuplement*

Les fonctions de peuplement comprennent la fonction de sélection (σ) et la fonction de jointure (⋈) ainsi que les fonctions de groupement, "*nest*", (η) et de dégroupement, "*unnest*" ($\eta^{-1}$) ; FP={σ, ⋈, η, $\eta^{-1}$}. Leur rôle est de déterminer l'extension des classes entrepôt en définissant des critères qualifiant les objets source pertinents à recopier dans l'entrepôt, ou bien, en regroupant des objets ou dégroupant des ensembles d'objets.

La fonction de **sélection** $\sigma_p$(cs) génère une classe entrepôt dont la structure est dérivée de celle de *cs* et dont l'extension est calculée à partir d'une restriction de l'extension source (exprimée au travers du prédicat de sélection *p*). La fonction de **jointure** $\bowtie_p(cs_1,cs_2)$ génère une classe entrepôt dont la structure est l'union de celles de $cs_1$ et $cs_2$ et dont l'extension est calculée en filtrant par le prédicat de jointure *p*, le produit cartésien entre les deux classes source $cs_1$ et $cs_2$.

La fonction **nest** $\eta_{\{p1,\ p2,\ldots,pn\}::attr}$(cs) et la fonction **unnest** $\eta^{-1}_{\{p1,\ p2,\ldots,pm\}}$(cs) permettent respectivement de grouper et de dégrouper les objets source pour peupler la classe entrepôt. Plus précisément, la fonction de regroupement transforme une classe source en créant une classe entrepôt telle que, pour chaque valeur des propriétés de $\{p_1, p_2,\ldots,p_n\}$, un ensemble de valeurs est construit (nommé *attr*) pour les autres propriétés. La fonction de dégroupement réalise l'opération contraire.

*Exemple* : Nous poursuivons l'exemple de la section 4.1 en ajoutant une sélection dans la définition de la fonction de construction appliquée à la classe source "*PRATICIEN*". L'administrateur définit la classe entrepôt "*Praticiens*" en sélectionnant l'ensemble des praticiens travaillant dans la région Midi-Pyrénées, puis en projetant les propriétés utiles aux décideurs ("*nom*", "*prenom*", "*annee_n*", "*specialite*", "*categorie*", "*ville*", "*densite*", "*departement*", "*consultations*"), augmentées d'un attribut calculé représentant le nombre d'enfants ("*nb_enfants*").

```
β^Praticien=π[o.nom,o.prenom,o.annee_n,o.categorie,o.specialite,o.PERSONNEadresse.ville,
           o.PERSONNEadresse.densite,o.PERSONNEadresse.departement,o.consultations]
               (o α[nb_enfants:count(p.enfants)](p ⋈[pp.travaille=c]
                     (pp PRATICIEN,
                      c σ [cc.adresse.region="Midi-Pyrenees"](cc CABINET))))
```

De plus, au moyen d'une fonction de construction plus complexe, l'administrateur extrait les informations concernant l'activité des praticiens stockées dans la source. Pour cela, il génère une classe "*Prescription*" qui est obtenue en regroupant les honoraires, les diagnostics et les prescripteurs de la classe intermédiaire calculée avec une jointure entre les visites effectuées chez les praticiens et les médicaments qu'ils prescrivent.

```
β^Prescription=η[g.honoraire,g.prescripteur,
             g.tension,g.poids,g.taille]::medicament
               (g π[vp.honoraire,vp.prescripteur,vp.tension,vp.poids,
                    vp.taille,vp.code,vp.generique,vp.categorie_molecule,
                    vp.type_molecule,vp.quantite,vp.tarif]
       (vp ⋈[m∈v.prescription]
```

```
(v ⋈[vv∊p.consultations](vv VISITE,
                          p ⋈[pp.travaille=c] (pp PRATICIEN,
                    c σ [cc.adresse.region="Midi-Pyrenees"](cc CABINET)),
  m MEDICAMENT)))
```

### 4.3 Fonctions ensemblistes

Les fonctions ensemblistes regroupent les fonctions traditionnelles de l'algèbre objet [SHAW90], c'est à dire l'union (∪), l'intersection (∩) et la différence (−) ; FE={∪, ∩, −}. Leur rôle est de combiner plusieurs classes sources afin de définir une nouvelle classe entrepôt.

### 4.4 Fonctions de hiérarchisation

Les précédentes fonctions génèrent des classes entrepôt dont la hiérarchie d'héritage n'est pas organisée ($Super^c=\emptyset$). Par conséquent, nous introduisons deux nouvelles fonctions visant à généraliser (Λ) et à spécialiser (Σ) les classes entrepôt afin de construire une hiérarchie d'héritage adaptée aux exigences spécifiques de l'entrepôt et de ses utilisateurs ; FH={Λ, Σ}.
La fonction de **généralisation** $\Lambda_{\{p1,\ p2,\ldots,ps\}}(c_1,c_2,\ldots,\ c_n)$ génère une super classe entrepôt à partir d'une ou plusieurs classes, en regroupant l'ensemble $\{p_1,\ p_2,\ldots,\ p_s\}$ des propriétés communes, spécifiées par l'administrateur.
La fonction de **spécialisation** $\Sigma_p(c_1,c_2,\ldots,\ c_n)$ génère une sous classe entrepôt à partir d'une ou plusieurs classes entrepôt $c_1,\ c_2,\ldots,\ c_n$.

<u>*Exemple*</u> : Nous complétons l'exemple de la section 4.2 dans lequel les classes entrepôt "*Praticien*" et "*Prescription*" ont été définies Afin d'organiser la hiérarchie d'héritage dans l'entrepôt, l'administrateur définit une super classe et une sous classe à la classe "*Praticien*", modélisant respectivement les personnes et les jeunes praticiens (dont l'année de naissance est supérieure à 1960).

```
β^Personne=Λ[o.nom, o.prenom, o.ville, o.densite, o.departement,
          o.annee_n, o.nb_enfants](o Praticien)
β^Jenue_Praticien=Σ[o.annee_n>1960](o Praticien)
```

## 5 Extraction des comportements

Le paradigme objet que nous adoptons, encapsule dans une même entité structure et comportement. Après avoir étudié l'extraction des données et de leur structure dans la section 4, cette section traite de la dérivation du comportement de ces données.
A notre connaissance, aucune proposition ne traite de ce problème dans le domaine des entrepôts ; la raison principale étant que les entrepôts de données sont habituellement développés dans un contexte relationnel où seules les structures sont dérivées par le biais de la technique des vues matérialisées [WIDO95] [GUPT95] [CHAU97].
Pour extraire le comportement dérivé des classes entrepôt, nous proposons un processus automatique, répondant aux critères suivants :

- Déterminer si les propriétés nécessaires à une méthode sont dérivées dans l'entrepôt,
- Déterminer si les objets manipulés par une méthode sont dérivées dans l'entrepôt,
- Déterminer si les méthodes indispensables à une autre méthode sont dérivées dans l'entrepôt.

### 5.1 Technique des matrices d'usage

Pour dériver les méthodes d'une classe source, nous devons déterminer si l'ensemble des éléments (propriétés, objets, méthodes) requis par chaque méthode est présent dans l'entrepôt. La solution que nous proposons s'inspire de la technique des matrices d'usage utilisées dans la conception des bases de données réparties [RAVA95]. Nous étendons le concept de matrice d'usage initialement proposé dans les bases de données réparties afin de l'adapter au contexte de la conception des entrepôts objets.
Dans notre contexte, une matrice d'usage (Figure 4) est définie de la manière suivante : pour chaque méthode (lignes de la matrice) la valeur 1 indique les critères (colonnes de la matrice) nécessaires au fonctionnement de la méthode. En outre, une ligne supplémentaire nommée "*Dérivé*" précise les critères présents (ou dérivés) dans l'entrepôt tandis qu'une colonne "*Dérivable*" permet de décider de l'extraction ou non de chaque méthode en fonction de la présence dans l'entrepôt des critères requis.

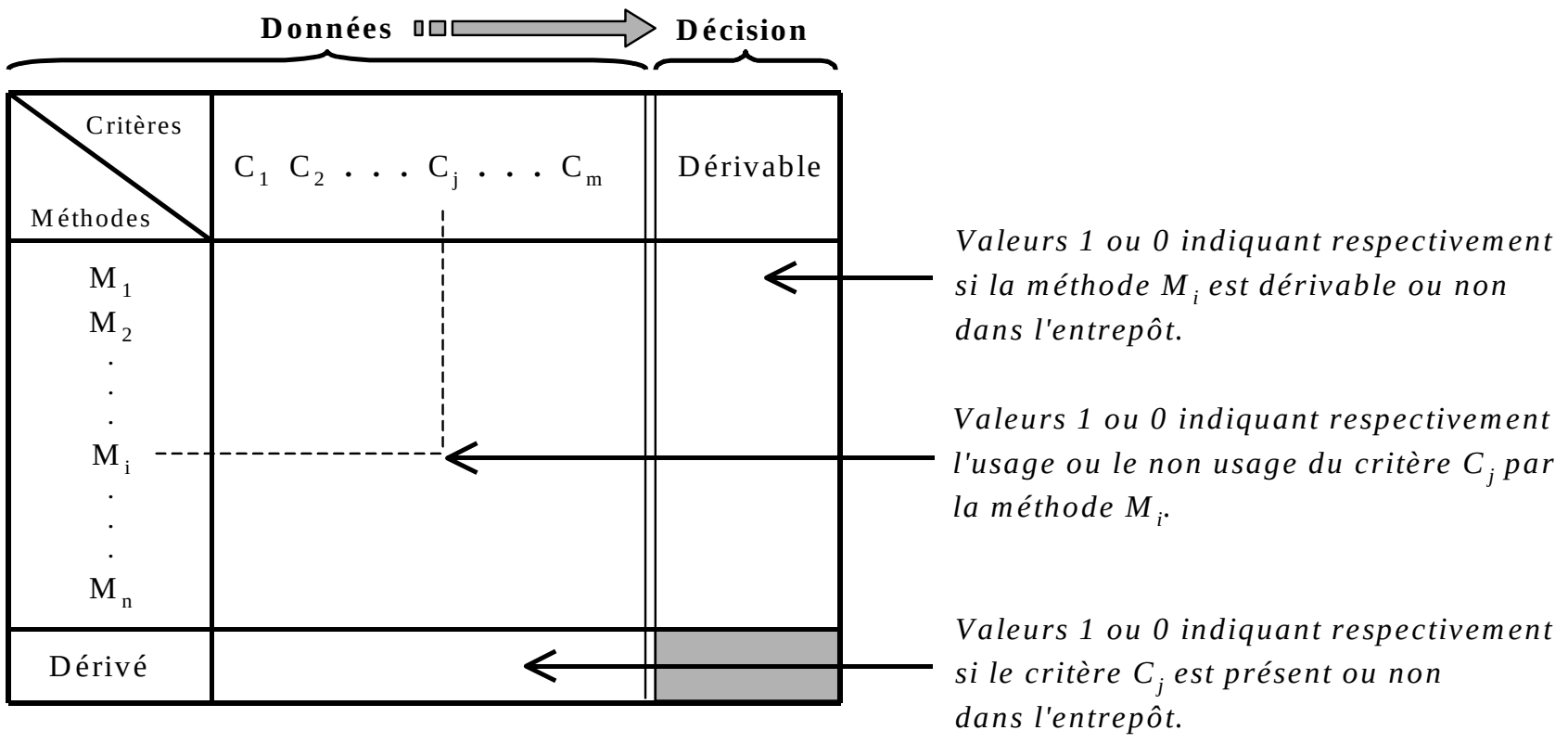

Figure 4 : *Principe des matrices d'usage.*

Nous utilisons trois matrices d'usage :

- la **matrice d'usage des propriétés** (MUP) se propose de déterminer si les propriétés nécessaires aux méthodes sont dérivées dans l'entrepôt.
- la **matrice d'usage des objets** (MUO) vise à déterminer si les objets manipulés par les méthodes sont présents dans l'entrepôt. En raison de la multitude des objets, il n'est pas réaliste de considérer les objets individuellement. En conséquence, nous définissons des groupes d'objets manipulés au travers de prédicats de sélection. Pour prendre en compte les prédicats de sélection des méthodes, nous utilisons la forme normale disjonctive ($E_1 \vee E_2 \vee \ldots \vee E_k$) qui permet la construction d'une disjonction d'éléments ayant une forte cohésion (chaque élément $E_i$ est une conjonction de prédicats simples $p_1 \wedge p_2 \wedge \ldots \wedge p_p$).
- la **matrice d'usage des méthodes** (MUM) sert à déterminer si les méthodes utilisées par chaque méthode sont disponibles dans l'entrepôt.

Une méthode n'accède qu'aux propriétés locales à la classe (les propriétés des autres classes sont manipulées au travers de méthodes d'accès) et seuls les objets de la classe sont manipulés par ses méthodes. Par contre, les méthodes d'une classe peuvent manipuler les méthodes locales à la classe mais également les méthodes publiques des autres classes.

Or, chaque fonction $Mapping^c$ génère une classe entrepôt dont il est nécessaire de définir le comportement dérivé de la source. Par conséquent, le processus d'extraction des comportements intervient à deux niveaux :

- **Localement** à chaque classe entrepôt générée. Une MUP et une MUO sont définies pour chaque classe entrepôt pour déterminer localement si toutes les propriétés nécessaires aux méthodes dérivables sont disponibles et si tous les objets manipulés sont dérivés.
- **Globalement** à l'entrepôt. Une MUM globale à tout l'entrepôt est construite pour déterminer si toutes les méthodes nécessaires aux méthodes dérivables sont disponibles.

Une méthode est dite dérivable si tous les éléments (propriétés, objets, autres méthodes) utilisés par la méthode sont disponibles.

## 5.2 Matrices d'usage locales

Cette partie décrit la phase de construction des matrices d'usage locales ainsi que leur analyse. La construction vise à générer la partie *Données* de la matrice tandis que l'analyse se propose de compléter la partie *Décision*, c'est à dire la colonne supplémentaire "*Dérivable*".

### 5.2.1 Construction des MU locales

Une MUP est construite pour chaque fonction de construction participant à l'élaboration d'un entrepôt. La phase de construction des MUP est réalisée lors de l'extraction des structures de données et du peuplement des classes entrepôt (section 4).

- Les lignes de la MUP correspondent aux méthodes des classes source impliquées dans la fonction de construction.
- Les colonnes de la MUP correspondent aux propriétés des classes source impliquées dans la fonction de construction.
- Chaque case (i,j) de la MUP contient la valeur 1 si la $i^{ième}$ méthode utilise la $j^{ième}$ propriété ; elle contient la valeur 0 sinon.
- La ligne supplémentaire "*Dérivé*" indique si la propriété est dérivée dans l'entrepôt.

<u>*Exemple*</u> : Prenons les exemples de fonctions d'extraction de la section 4. L'élaboration des structures de données de l'entrepôt et le peuplement de l'entrepôt s'appuient notamment sur deux fonctions de construction permettant de générer par extraction les classes entrepôt "*Praticien*" et "*Prescription*".

| Opérations \ Propriétés | nom | prenom | annee_n | PERSONNEadresse.libelle | PERSONNEadresse.code | PERSONNEadresse.ville | PERSONNEadresse.departement | PERSONNEadresse.region | PERSONNEadresse.densite | mrie | enfants | parents | num_prat | categorie | specialite | type_convention | diplomes | travaille | consultations | intitule | CABINETadresse.libelle | CABINETadresse.code | CABINETadresse.ville | CABINETadresse.departement | CABINETadresse.region | CABINETadresse.densite | membres | **Dérivable** |
|---|---|---|---|---|---|---|---|---|---|---|---|---|---|---|---|---|---|---|---|---|---|---|---|---|---|---|---|---|
| age | 0 | 0 | 1 | 0 | 0 | 0 | 0 | 0 | 0 | 0 | 0 | 0 | 0 | 0 | 0 | 0 | 0 | 0 | 0 | 0 | 0 | 0 | 0 | 0 | 0 | 0 | 0 | ? |
| est_rural | 0 | 0 | 0 | 0 | 0 | 1 | 0 | 0 | 0 | 0 | 0 | 0 | 0 | 0 | 0 | 0 | 0 | 0 | 0 | 0 | 0 | 0 | 0 | 0 | 0 | 0 | 0 | ? |
| est_urbain | 0 | 0 | 0 | 0 | 0 | 1 | 0 | 0 | 0 | 0 | 0 | 0 | 0 | 0 | 0 | 0 | 0 | 0 | 0 | 0 | 0 | 0 | 0 | 0 | 0 | 0 | 0 | ? |
| est_interne | 0 | 0 | 0 | 0 | 0 | 0 | 0 | 0 | 0 | 0 | 0 | 0 | 0 | 1 | 0 | 0 | 0 | 0 | 0 | 0 | 0 | 0 | 0 | 0 | 0 | 0 | 0 | ? |
| est_generaliste | 0 | 0 | 0 | 0 | 0 | 0 | 0 | 0 | 0 | 0 | 0 | 0 | 0 | 1 | 0 | 0 | 0 | 0 | 0 | 0 | 0 | 0 | 0 | 0 | 0 | 0 | 0 | ? |
| taux_remb | 0 | 0 | 0 | 0 | 0 | 0 | 0 | 0 | 0 | 0 | 0 | 0 | 0 | 0 | 0 | 1 | 0 | 0 | 0 | 0 | 0 | 0 | 0 | 0 | 0 | 0 | 0 | ? |
| **Dérivé** | 1 | 1 | 1 | 0 | 0 | 1 | 1 | 0 | 1 | 0 | 0 | 0 | 0 | 1 | 1 | 0 | 0 | 0 | 1 | 0 | 0 | 0 | 0 | 0 | 0 | 0 | 0 | |

**Figure 5** : Construction de la MUP servant à l'extraction du comportement de "*Praticien*".

L'expression suivante représente la fonction de construction utilisée pour générer la classe entrepôt "*Praticien*" : $f_4$ ($f_3$ ($f_2$ ($f_1$(CABINET), PRATICIEN))) avec $f_4 \equiv \pi$, $f_3 \equiv \alpha$, $f_2 \equiv \bowtie$, $f_1 \equiv \sigma$. Une MUP est donc construite pour déterminer le comportement de "*Praticien*". Elle est constituée de la manière suivante :

- Chaque ligne correspond aux méthodes des classes source impliquées dans la fonction de construction (PRATICIEN, PERSONNE, CABINET). La classe PERSONNE participe à la construction de la matrice d'usage puisque cette classe est super classe de PRATICIEN.
- Chaque colonne correspond aux propriétés des classes source impliquées. Notons que certaines propriétés sont renommées en cas de redondance dans plusieurs classes ; par exemple l'attribut adresse est renommé en plaçant en préfixe le nom de la classe qui le contient puisque cet attribut apparaît à la fois dans PERSONNE et CABINET.
- Chaque case est complétée en s'appuyant sur une analyse du corps des méthodes dont le résultat est conservé dans les méta-données. Par exemple, la méthode "*age()*" de la classe PERSONNE utilise l'attribut "*annee_n*" ; l'intersection de la ligne correspondant à la méthode "*age()*" et de la colonne correspondant à la propriété "*annee_n*" prend la valeur 1 tandis que les autres cases de la ligne sont fixées à 0.
- La ligne supplémentaire "*Dérivé*" est construite automatiquement à partir de la fonction de construction. Ainsi $f_4 \equiv \pi$ détermine que les propriétés "*nom*", "*prenom*", "*annee_n*", "*PERSONNEadresse.ville*","*categorie*", "*specialite*", "*PERSONNEadresse.departement*", "*PERSONNEadresse.densite*", "*consultation*" sont dérivées dans l'entrepôt ; les cases d'intersection sur la ligne supplémentaire "*Dérivé*" reçoivent la valeur 1 tandis que les autres cases sont positionnées à la valeur 0.

L'expression suivante représente la fonction de construction utilisée pour générer la classe entrepôt "*Prescription*" : $f_8$ ($f_7$ ($f_6$ ($f_5$ ($f_2$ ($f_1$(CABINET), PRATICIEN), VISITE), MEDICAMENT))) avec $f_8 \equiv \eta$, $f_7 \equiv \pi$, $f_6 \equiv \eta$, $f_5 \equiv \bowtie$, $f_2 \equiv \bowtie$, $f_1 \equiv \sigma$. Une MUP est construite pour déterminer le comportement de la classe entrepôt "*Prescription*".

Cependant, il est possible d'optimiser cette matrice : il suffit de supprimer les colonnes de la matrice contenant uniquement des 0. En effet, cela signifie que la propriété n'est utilisée par aucune méthode dérivable et que la propriété n'est pas dérivée. Par conséquence, cette colonne est inutile. La Figure 6 décrit la MUP optimisée pour la classe entrepôt "*Prescription*".

En outre, nous constatons que la propriété "*prescription*", qui n'est pas projetée par la fonction de construction ($Mapping^{Prescription}$), est néanmoins considérée comme dérivée dans l'entrepôt (la ligne "*Dérivé*" contient la valeur 1).

| Propriétés / Méthodes | honoraire | symptomes | tension.max | tension.min | poids | taille | prescription | prescripteur | code | generique | categorie_molecule | type_molecule | quantite | tarif | taux_secu | **Dérivable** |
|---|---|---|---|---|---|---|---|---|---|---|---|---|---|---|---|---|
| montant_euro | 0 | 0 | 0 | 0 | 0 | 0 | 0 | 0 | 0 | 0 | 0 | 0 | 0 | 0 | 0 | ? |
| montant_prescrit | 0 | 0 | 0 | 0 | 0 | 0 | 1 | 0 | 0 | 0 | 0 | 0 | 0 | 0 | 0 | ? |
| nb_symptomes | 0 | 1 | 0 | 0 | 0 | 0 | 0 | 0 | 0 | 0 | 0 | 0 | 0 | 0 | 0 | ? |
| affiche_tension | 0 | 0 | 1 | 1 | 0 | 0 | 0 | 0 | 0 | 0 | 0 | 0 | 0 | 0 | 0 | ? |
| est_obese | 0 | 0 | 0 | 0 | 1 | 1 | 0 | 0 | 0 | 0 | 0 | 0 | 0 | 0 | 0 | ? |
| cout_secu | 1 | 0 | 0 | 0 | 0 | 0 | 1 | 0 | 0 | 0 | 0 | 0 | 0 | 0 | 0 | ? |
| montant_remb | 0 | 0 | 0 | 0 | 0 | 0 | 0 | 0 | 0 | 0 | 0 | 0 | 1 | 1 | 1 | ? |
| **Dérivé** | 1 | 0 | 1 | 1 | 1 | 1 | 1 | 1 | 1 | 1 | 1 | 1 | 1 | 1 | 0 | |

**Figure 6 :** Construction de la MUP optimisée pour l'extraction du comportement de "*Prescription*".

En effet, cette relation relie les classes VISITE et MEDICAMENT qui sont fusionnées dans la fonction de construction par une fonction de jointure ; par conséquence, les méthodes des deux classes peuvent accéder aux propriétés de la classe liée sans utiliser la relation.

De manière analogue, il est possible de construire des MUO pour identifier les objets manipulés par les méthodes. Remarquons qu'une méthode est applicable sur tous les objets de la classe, mais le traitement de la méthode peut ne concerner qu'une partie des objets. Contrairement à la MUP, la MUO n'est pas indispensable pour le bon fonctionnement de l'entrepôt.

### 5.2.2 Analyse des MU locales

L'analyse consiste à compléter la colonne supplémentaire "*Dérivable*" des matrices locales. Pour chaque méthode de la matrice locale, le processus d'analyse identifie les éléments (propriétés ou prédicats d'objet) utilisés.

- Si tous les éléments utiles à la méthode sont dérivés dans l'entrepôt, la case de la colonne "*Dérivable*" reçoit la valeur 1,
- Sinon la case de la colonne "*Dérivable*" reçoit la valeur 0.

Le processus d'analyse des MU locales est défini par la Figure 7. L'algorithme proposé réalise l'analyse d'une ligne i (c'est à dire une méthode) dans une MU locale. Elle détermine si l'ensemble des propriétés (pour une MUP) ou des objets (pour une MUO) est disponible (dérivé) dans l'entrepôt. Dans ce cas, la case de la colonne "*Dérivable*" prend la valeur 1, sinon elle prend la valeur 0 et l'ensemble des éléments manquants est retourné en sortie.

```
ALGORITHME. Analyse_Locale(i,MU)
Entrées :
   i  : indice identifiant la méthode analysée dans la matrice
   MU : matrice d'usage locale analysée
Sortie :
   ensemble des éléments (propriétés ou prédicats d'objets) manquantes utilisées par la méthodes i
Début
ens←∅;
pour j←1 à colonnes(MU) faire
   si MU[i,j]=1 alors
```

```
      si MU[lignes(MU)+1,j]=0 alors ens←ens∪{j};
si (ens=∅) alors MU[i,colonnes(MU)+1]←1;
sinon MU[i,colonnes(MU)+1]←0;
retourne ens;
Fin.
```

**Figure 7 :** Algorithme de l'analyse d'une MU locale.

Remarquons que les fonctions *lignes(MU)* et *colonnes(MU)* retournent respectivement le nombre de lignes et le nombre de colonnes composant la matrice d'usage MU.

<u>*Exemple*</u> : Nous poursuivons l'exemple précédent. Le processus d'analyse peut être appliqué sur chaque ligne de la matrice (c'est à dire pour chaque méthode potentiellement extractible) ; le processus d'analyse de la MUP complète la colonne "*Dérivable*". La Figure 8 décrit la matrice obtenue.

- La MUP associée à "*Praticien*" décide que les méthodes "*age()*", "*est_interne()*" et "*est_generaliste()*" disposent dans la classe entrepôt "*Praticien*" des propriétés nécessaires à leur fonctionnement.
- La MUP associée à "*Prescription*" décide que les méthodes "*montant_euro()*", "*montant_prescrit()*", "*affiche_tension()*", "*est_obese()*" et "*cout_secu()*" disposent dans la classe entrepôt "*Prescription*" des propriétés nécessaires à leur fonctionnement.

| Méthodes \ Propriétés | nom | prenom | annee_n | PERSONNEadresse.libelle | PERSONNEadresse.code | PERSONNEadresse.ville | PERSONNEadresse.departement | PERSONNEadresse.region | PERSONNEadresse.densite | marie | enfants | parents | num_prat | categorie | specialite | type_convention | diplomes | travaille | consultations | intitule | CABINETadresse.libelle | CABINETadresse.code | CABINETadresse.ville | CABINETadresse.departement | CABINETadresse.region | CABINETadresse.densite | membres | **Dérivable** |
|---|---|---|---|---|---|---|---|---|---|---|---|---|---|---|---|---|---|---|---|---|---|---|---|---|---|---|---|---|
| age | 0 | 0 | 1 | 0 | 0 | 0 | 0 | 0 | 0 | 0 | 0 | 0 | 0 | 0 | 0 | 0 | 0 | 0 | 0 | 0 | 0 | 0 | 0 | 0 | 0 | 0 | 0 | 1 |
| est_rural | 0 | 0 | 0 | 0 | 1 | 0 | 0 | 0 | 0 | 0 | 0 | 0 | 0 | 0 | 0 | 0 | 0 | 0 | 0 | 0 | 0 | 0 | 0 | 0 | 0 | 0 | 0 | 0 |
| est_urbain | 0 | 0 | 0 | 0 | 1 | 0 | 0 | 0 | 0 | 0 | 0 | 0 | 0 | 0 | 0 | 0 | 0 | 0 | 0 | 0 | 0 | 0 | 0 | 0 | 0 | 0 | 0 | 0 |
| est_interne | 0 | 0 | 0 | 0 | 0 | 0 | 0 | 0 | 0 | 0 | 0 | 0 | 0 | 1 | 0 | 0 | 0 | 0 | 0 | 0 | 0 | 0 | 0 | 0 | 0 | 0 | 0 | 1 |
| est_generaliste | 0 | 0 | 0 | 0 | 0 | 0 | 0 | 0 | 0 | 0 | 0 | 0 | 0 | 1 | 0 | 0 | 0 | 0 | 0 | 0 | 0 | 0 | 0 | 0 | 0 | 0 | 0 | 1 |
| taux_remb | 0 | 0 | 0 | 0 | 0 | 0 | 0 | 0 | 0 | 0 | 0 | 0 | 0 | 0 | 0 | 1 | 0 | 0 | 0 | 0 | 0 | 0 | 0 | 0 | 0 | 0 | 0 | 0 |
| **Dérivé** | 1 | 1 | 1 | 0 | 0 | 1 | 1 | 0 | 1 | 0 | 0 | 0 | 0 | 1 | 1 | 0 | 0 | 0 | 1 | 0 | 0 | 0 | 0 | 0 | 0 | 0 | 0 | |

| Méthodes \ Propriétés | honoraire | symptomes | tension.max | tension.min | poids | taille | prescription | prescripteur | code | generique | categorie_molecule | type_molecule | quantite | tarif | taux_secu | **Dérivable** |
|---|---|---|---|---|---|---|---|---|---|---|---|---|---|---|---|---|
| montant_euro | 0 | 0 | 0 | 0 | 0 | 0 | 0 | 0 | 0 | 0 | 0 | 0 | 0 | 0 | 0 | 1 |
| montant_prescrit | 0 | 0 | 0 | 0 | 0 | 0 | 1 | 0 | 0 | 0 | 0 | 0 | 0 | 0 | 0 | 0 |
| nb_symptomes | 0 | 1 | 0 | 0 | 0 | 0 | 0 | 0 | 0 | 0 | 0 | 0 | 0 | 0 | 0 | 0 |
| affiche_tension | 0 | 0 | 1 | 1 | 0 | 0 | 0 | 0 | 0 | 0 | 0 | 0 | 0 | 0 | 0 | 1 |
| est_obese | 0 | 0 | 0 | 0 | 1 | 1 | 0 | 0 | 0 | 0 | 0 | 0 | 0 | 0 | 0 | 1 |
| cout_secu | 1 | 0 | 0 | 0 | 0 | 0 | 1 | 0 | 0 | 0 | 0 | 0 | 0 | 0 | 0 | 1 |
| montant_remb | 0 | 0 | 0 | 0 | 0 | 0 | 0 | 0 | 0 | 0 | 0 | 0 | 1 | 1 | 1 | 0 |
| **Dérivé** | 1 | 0 | 1 | 1 | 1 | 1 | 1 | 1 | 1 | 1 | 1 | 1 | 1 | 1 | 0 | |

**Figure 8 :** Analyse des MUP associées à "*Praticien*" et "*Prescription*".

## 5.3 Matrice d'usage globale

Cette partie décrit la phase de construction de la MUM ainsi que son analyse pour compléter la colonne supplémentaire qui permet de décider de l'extraction ou non d'une méthode.

### 5.3.1 Construction de la MU globale

Une seule MUM est construite pour l'entrepôt. Cette matrice est élaborée de la manière suivante :

- Chaque ligne de la MUM correspond à une méthode des classes source impliquées dans les fonctions de construction (utilisées pour générer les structures de l'entrepôt).
- Chaque colonne de la MUM correspond également à une méthode des classes source impliquées dans les fonctions de construction.
- Chaque case (i,j) de la MUM contient la valeur 1 si la $i^{ième}$ méthode utilise la $j^{ième}$ méthode sinon elle contient la valeur 0.

La ligne supplémentaire est fixée à –1 tant que la méthode n'a pas été analysée puisqu'il n'est pas possible à ce stade de savoir si la méthode est dérivée. Cependant, les matrices d'usage locales peuvent indiquer si certaines méthodes sont non dérivables (car des propriétés nécessaires sont absentes).

*Exemple* : Reprenons les exemples de fonctions d'extraction de la section 4. La Figure 9 décrit la matrice globale. Elle est constituée des méthodes des classes source impliquées dans l'élaboration de l'entrepôt (PERSONNE, CABINET, PRATICIEN, VISITE et MEDICAMENT).
La méthode "*cout_secu*" utilise la méthode "*montant_remb*" qui n'est pas dérivable dans l'entrepôt car cette dernière ne dispose pas de toutes les propriétés qui lui sont nécessaires ("*taux_secu*") ; se reporter à la Figure 8.

| Méthodes \ Méthodes | age | est_rural | est_urbain | est_interne | est_generaliste | taux_remb | montant_euro | montant_prescrit | nb_symptomes | affiche_tension | est_obese | cout_secu | montant_remb | **Dérivable** |
|---|---|---|---|---|---|---|---|---|---|---|---|---|---|---|
| age | 0 | 0 | 0 | 0 | 0 | 0 | 0 | 0 | 0 | 0 | 0 | 0 | 0 | 1 |
| est_rural | 0 | 0 | 0 | 0 | 0 | 0 | 0 | 0 | 0 | 0 | 0 | 0 | 0 | 1 |
| est_urbain | 0 | 0 | 0 | 0 | 0 | 0 | 0 | 0 | 0 | 0 | 0 | 0 | 0 | 1 |
| est_interne | 0 | 0 | 0 | 0 | 0 | 0 | 0 | 0 | 0 | 0 | 0 | 0 | 0 | 1 |
| est_generaliste | 0 | 0 | 0 | 0 | 0 | 0 | 0 | 0 | 0 | 0 | 0 | 0 | 0 | 1 |
| taux_remb | 0 | 0 | 0 | 0 | 0 | 0 | 0 | 0 | 0 | 0 | 0 | 0 | 0 | 1 |
| montant_euro | 0 | 0 | 0 | 0 | 0 | 0 | 0 | 1 | 0 | 0 | 0 | 0 | 0 | 1 |
| montant_prescrit | 0 | 0 | 0 | 0 | 0 | 0 | 0 | 0 | 0 | 0 | 0 | 0 | 0 | 1 |
| nb_symptomes | 0 | 0 | 0 | 0 | 0 | 0 | 0 | 0 | 0 | 0 | 0 | 0 | 0 | 1 |
| affiche_tension | 0 | 0 | 0 | 0 | 0 | 0 | 0 | 0 | 0 | 0 | 0 | 0 | 0 | 1 |
| est_obese | 0 | 0 | 0 | 0 | 0 | 0 | 0 | 0 | 0 | 0 | 0 | 0 | 0 | 1 |
| cout_secu | 0 | 0 | 0 | 0 | 0 | 0 | 0 | 0 | 0 | 0 | 0 | 0 | 1 | 0 |
| montant_remb | 0 | 0 | 0 | 0 | 0 | 0 | 0 | 0 | 0 | 0 | 0 | 0 | 0 | 1 |
| **Dérivé** | -1 | -1 | -1 | -1 | -1 | -1 | -1 | -1 | -1 | -1 | -1 | -1 | 0 | |

**Figure 9 :** Exemple d'une MUM.

### *5.3.2 Analyse de la MU globale*

Le processus d'analyse par la MUM est plus complexe que le précédent. Cela réside dans le fait que l'on ne peut identifier au préalable les méthodes manipulées qui sont présentes dans l'entrepôt. Une difficulté supplémentaire est liée au fait qu'une méthode i manipulant une autre méthode i', peut être manipulée elle même par cette méthode i'. Ce cas particulier d'appel mutuel entre méthodes génère alors un cycle dans le processus d'analyse. Dans ce cas, l'intervention de l'administrateur est demandée pour indiquer si les méthodes sont dérivables.
Le processus d'analyse de la MU globale est défini par la Figure 10. L'algorithme proposé réalise l'analyse d'une ligne i (c'est à dire une méthode) dans une MU globale. Elle détermine pour chaque méthode manipulée j par la méthode i si cette méthode j est présente dans l'entrepôt. Si j n'est pas encore analysée complètement, le processus lance son analyse. Après l'analyse complète, si l'ensemble des méthodes nécessaires est dérivé, la case de la colonne "*Dérivable*" prend la valeur 1, sinon elle prend la valeur 0 et l'ensemble des méthodes manquantes est retourné en sortie.

```
ALGORITHME. Analyse_Globale(i,MUM,visite)
Entrées :
   i      : indice identifiant la méthode analysée dans la matrice
   visite : vecteur détectant les cycles
Sortie :
   ensemble des méthodes manquantes utilisées par la méthodes i
Début
ens←∅;
pour j←1 à colonnes(MUM) faire
   début
   visite[i]←1;
   si MUM[i,j]=1 alors
      cas MUM[lignes(MUM)+1,j]=0 :  début // méthode manquante
         ens←ens∪{j};
         fin;
      cas MUM[lignes(MUM)+1,j]=-1 : début // méthode non analysée
         si visite[j]=1 alors ens←ens∪{j}; // détection d'un cycle
         sinon début // analyse
               visite[i]←1;
               MUP←determineMUP(j,MUM);
               MUO←determineMUO(j,MUM);
               jj←numero(j,MUM,MUP); // equivalent à numero(j,MUM,MUO)
               ens=ens∪Analyse_Locale(jj,MUP);
               ens=ens∪Analyse_Locale(jj,MUO);
               ens=ens∪Analyse_Globale(j,MUM,visite);
               fin;
         fin;
   fin;
```

```
si (ens=∅) alors MUM[i,colonnes(MUM)+1]←1;
sinon MUM[i,colonnes(MUM)+1]←0;
retourne ens;
Fin.
```

**Figure 10 :** Algorithme de l'analyse de la MU globale.

Dans le cas d'un appel récursif ("*MUM[m+1,j]=-1*" et "*visite[j]=0*"), les fonctions *determineMUP*(*num_methode*,*MU_globale*) et *determineMUO*(*num_methode*,*MU_globale*) retournent la MUP et la MUO associées à la classe de la méthode *num_méthode*. La fonction *numero*(*num_methode*,*MU_globale*,*MU_locale*) retourne le numéro local à la matrice *MU_locale* correspondant à la méthode *num_methode* dans la *MU_globale.*

## 5.4 Dérivation du comportement

En s'appuyant sur l'analyse des matrices d'usage locales (MUP, MUO) et de la matrice d'usage globale (MUM), nous déterminons l'ensemble des méthodes dérivables.

Lorsqu'une méthode n'est pas dérivée, le processus indique les éléments (propriétés, objets méthodes) nécessaires manquants, afin que l'administrateur puisse ajuster l'entrepôt à ses besoins en dérivant les éléments nécessaires à une méthode qu'il souhaite extraire.

Le processus de dérivation du comportement des classes entrepôt est défini par la Figure 11. Les analyses des matrices locales sont réalisées puis l'analyse de la matrice globale est effectuée. A chaque méthode analysée localement ou globalement, l'ensemble des éléments manipulés manquants est affiché à l'administrateur. Lorsque les analyses sont terminées, il est possible d'indiquer le comportement dérivé des classes entrepôt.

```
ALGORITHME. Deriver_Comportement()
Sortie :
   ensemble des méthodes dérivables
Début
// analyses locales (MUP, MUO)
pour chaque MUP faire
   pour i←1 à colonnes(MUP) faire
      si ((manque←Analyse_Locale(i,MUP))≠∅) alors affiche(i,manque);
pour chaque MUO faire
   pour i←1 à colonnes(MUO) faire
      si ((manque←Analyse_Locale(i,MUO))≠∅) alors affiche(i,manque);
// analyse globale (MUM)
pour i←1 à colonnes(MUM) faire
   début
   pour ii←1 à colonnes(MUM) faire visite[ii]←0;
   si ((manque←Analyse_Globale(i,MUM,visite))≠∅) alors affiche(i,manque);
   fin;
// décision comportement dérivable
res←∅;
pour i←1 à colonnes(MUM) faire
   début
   MUP←determineMUP(i,MUM);
   MUO←determineMUO(i,MUM);
   ii←numero(i,MUM,MUP); // equivalent à numero(i,MUM,MUO)
   si MUP[ii,colonnes(MUP)+1]=1 &&
      MUO[ii,colonnes(MUO)+1]=1 &&
      MUM[i,colonnes(MUM)+1]=1 alors res←res∪{i};
      fin;
retourner res;
Fin.
```

**Figure 11 :** Algorithme d'extraction des comportements des classes entrepôt.

## 6 Conclusion

Cet article traite de la conception des entrepôts de données complexes. Nous avons d'abord défini un modèle d'entrepôt de données complexes et temporelles, basé sur le paradigme objet. Les principales caractéristiques de notre modèle sont les suivantes :

- Le concept d'objet entrepôt modélise l'état courant d'une information extraite, ainsi que des états passés (représentant les évolutions de l'objet sous une forme détaillée) et des états archivés (correspondant aux évolutions de l'objet décrites sous une forme résumée). L'intérêt de cette modélisation est de conserver les données de l'entrepôt ainsi que leurs évolutions à un niveau de détail pertinent, limitant le stockage des évolutions.
- Le concept de classe entrepôt intègre les caractéristiques de notre approche par une fonction de construction, un filtre temporel et un filtre d'archives.
- Le concept d'environnement permet de définir simplement les parties temporelles homogènes (même période de rafraîchissement, même critère d'archivage,…).

Dans un second temps, nous avons présenté les processus d'extraction nécessaires pour l'élaboration des classes entrepôt. Ce processus intègre l'aspect statique des données (données et structures des données) ainsi que l'aspect dynamique (comportement des données).

- L'extraction des structures et des données correspondantes (pour générer les structures et les objets des classes entrepôt) est réalisée au travers d'une fonction de construction, composée de fonctions de structuration, ensemblistes, de peuplement et/ou de hiérarchisation.
- L'extraction du comportement des données (pour générer celui des classes entrepôt) s'appuie sur un processus d'analyse des éléments nécessaires aux méthodes (propriétés, objets et méthodes). Pour cela, nous avons défini trois critères d'analyse au travers des matrices d'usage des propriétés, des objets et des méthodes. Le processus calcule automatiquement les méthodes dérivables et indique à l'administrateur les éléments manquants des méthodes non dérivées.

Notre étude a fait l'objet d'un développement au travers du prototype GEDOOH[3] [BRET99] [RAVA99], acronyme de <u>G</u>énérateur d'<u>E</u>ntrepôts de <u>D</u>onnées <u>O</u>rientées <u>O</u>bjet et <u>H</u>istorisées, comportant une interface (visualisant graphiquement la source globale et l'entrepôt de données) et un module générateur (permettant de créer automatiquement des entrepôts ainsi que les processus d'alimentation et de rafraîchissement). GEDOOH est opérationnel : il comprend 5000 lignes de code Java (jdk1.2) pour l'interface et 1500 lignes de code O2C pour le paquetage de classes, implanté au dessus du SGBD hôte utilisé (O2).

Actuellement, nous étudions l'extension du langage d'interrogation OQL afin de prendre en compte les caractéristiques des objets entrepôt (manipulation du temps, opérations manipulant des séries temporelles d'états…). D'autre part, il est souhaitable de mener une étude concernant une méthode de conception de l'entrepôt de données ainsi que des magasins de données. Actuellement, seul [GOLF99] propose une démarche de construction d'un magasin multidimensionnel à partir d'un schéma E/A. Dans notre contexte objet, nous élaborons une méthode (modèles et formalismes, démarche) $O_2$MDD [BRET00] destinée à la construction de bases de données décisionnelles multidimensionnelles et orientées objet.

## 7 Références

---

[3] *http://www.irit.fr/SSI/ACTIVITES/EQ_SIG/gedooh.html*

# 8 Annexe : Schéma d'une source globale

Nous décrivons un exemple complet, issu d'une application médicale et simplifié dans un souci de clarté. Notre exemple comprend :

- trois sources de données inspirées par des bases opérationnelles médicales,
- une source globale intégrant l'ensemble des informations contenues dans ces différentes sources,
- un entrepôt collectant les informations relatives aux praticiens de la région Midi-Pyrénées dont le processus d'extraction des données, des structures de données et des comportements sert d'illustration aux sections 4 et 5,
- un magasin de données supportant des analyses concernant les honoraires des praticiens en fonction de leur répartition dans la région Midi-Pyrénées.

## *8.1 Sources de données*

- **Source 1 :** (Relationnelle)

Cette source est décrite en relationnel (Oracle) et contient des informations concernant les praticiens de santé ; il s'agit d'un extrait d'une base opérationnelle de l'Assurance Maladie.

```
create table PRATICIEN(num_prat varchar2(17), nom_prat varchar2(20),
                       prenom_prat varchar2(20), categorie varchar2(30),
                       specialite varchar2(30));
create table CABINET(intitule varchar2(25), libelle_adr varchar2(50),
                     code_adr number, ville varchar(30), num_prat varchar2(17));
```

- **Source 2 :** (Fichier Texte)

Cette source décrit la structure d'un fichier texte, obtenu par extraction et filtrage de documents Web (disponibles sur le site de l'INSEE). Cette source regroupe des informations sur l'organisation géographique de la France.

```
Colonne1 → NomRegion, Colonne2 → NomDepartement, Colonne3 → NomVille, Colonne4 → densite
```

- **Source 3 :** (Objet)

Cette source est décrite en ODL (O2) et contient des informations concernant les patients qui consultent des praticiens. Cette source correspond à une base de données dans des hôpitaux.

```
public class PERSONNE
type tuple (
  nom:string,
  prenom:string,
  annee_n:integer,
  adresse:tuple(
        libelle:string,
        code:integer,
        ville:string),
  marie:PERSONNE,
  enfants:set(PERSONNE),
  parents:set(PERSONNE)  )
method public age():integer,
       public est_rural():boolean,
       public est_urbain():boolean
end;
public class PATIENT
inherit PERSONNE
type tuple (
  num_secu:string,
  cle_secu:string,
  visites:list(VISITE)  )
end;
public class PRATICIEN
inherit PERSONNE
type tuple (
  num_prat:string,
  categorie:string,
  specialite:string,
  type_convention:integer,
  diplomes:set(string)  )
method public est_interne():boolean,
       public est_generaliste():boolean,
       public taux_remb():real
end;
public class VISITE
type tuple (
  honoraire:real,
  symptomes:list(string),
  tension:tuple(
            max:integer,
            min:integer),
  poids:real,
  taille:real,
  temperature:real,
  diagnostic:string,
  prescription:set(MEDICAMENT),
  patient:PATIENT,
  prescripteur:PRATICIEN  )
method public montant_euro():real,
       public montant_prescrit():real,
       public nb_symptomes():integer,
       public affiche_tension(),
       public est_obese():boolean,
       public cout_secu():real,
end;
public class MEDICAMENT
type tuple (
  code:string,
  generique:boolean,
  categorie_molecule:string,
  type_molecule:string,
```

```
  posologie:string,
  quantite:integer,
  tarif:real,
  taux_secu:real,
  fabriquant:string  )
method public montant_remb():real
end;
```

## 8.2 Source globale

Pour décrire la source globale, nous utilisons le modèle objet standard : modèle de l'ODMG [CATT95].

- **Source Globale** : (ODMG)

```
interface PERSONNE {
  attribute String nom;
  attribute String prenom;
  attribute Short annee_n;
  attribute Struct T_Adresse{
        String libelle,
        String code,
        String ville,
        String departement,
        String region,
        Short densite} adresse;
  relationship PERSONNE marie
            inverse PERSONNE::marie;
  relationship Set<PERSONNE> enfants
            inverse PERSONNE::parents;
  relationship Set<PERSONNE> parents
            inverse PERSONNE::enfants;
  Short age();
  Boolean est_rural();
  Boolean est_urbain(); }
interface PATIENT (extend PERSONNE) {
  attribute String num_secu;
  attribute String cle_secu;
  relationship List<VISITE> visites
             inverse PATIENT::patient; }
interface PRATICIEN (extend PERSONNE) {
  attribute String num_prat;
  attribute String categorie;
  attribute String specialite;
  attribute Double type_convention;
  attribute Set<String> diplomes;
  relationship CABINET travaille
           inverse CABINET::membres;
  relationship List<VISITE> consultations
           inverse VISITE::prescripteur;
  Boolean est_interne();
  Boolean est_generaliste();
  Double taux_remb(); }
interface CABINET {
  attribute String intitule;
  attribute Struct T_Adresse{
        String libelle,
        String code,
        String ville,
        String departement,
        String region,
        Short densite} adresse;
  relationship Set<PRATICIEN> membres
             inverse PRATICIEN::travaille; }
interface VISITE {
  attribute Double honoraire;
  attribute List<String> symptomes;
  attribute Struct T_Tension{
        Short max,
        Short min} tension;
  attribute Double poids;
  attribute Double taille;
  attribute Double temperature;
  attribute String diagnostic;
  relationship Set<MEDICAMENT> prescription;
  relationship PATIENT patient
          inverse PATIENT::visites;
  relationship PRATICIEN prescripteur
          inverse PRATICIEN::consultations;
  Double montant_euro();
  Double montant_prescrit();
  Short nb_sympotmes();
  String affiche_tension();
  Boolean est_obese();
  Double cout_secu(); }
interface MEDICAMENT {
  attribute String code;
  attribute Boolean generique;
  attribute String categorie_molecule;
  attribute String type_molecule;
  attribute String posologie;
  attribute Short quantite;
  attribute Double tarif;
  attribute Double taux_secu;
  attribute String fabriquant;
  Double montant_remb(); }
```

## 8.3 Entrepôt de données

Prenons le cas où des médecins conseils de la sécurité sociale (décideurs) souhaitent analyser l'activité des praticiens travaillant dans la région Midi-Pyrénées. L'administrateur crée un entrepôt contenant l'information utile aux décideurs.

- **Entrepôt de données** :

```
interface Personne {
  attribute String nom;
  attribute String prenom;
  attribute Short annee_n;
  attribute String ville;
  attribute Short densite;
  attribute String departement;
  attribute Short nb_enfants;
  Short age();  }
interface Praticien (extend Personne) {
  attribute String categorie;
  attribute String specialite;
  relationship List<Prescription>
                                 consultations
        inverse Prescription::prescripteur;
  Boolean est_interne();
  Boolean est_generaliste();  }
interface Jeune_Praticien
                             (extend Praticien) { }
interface Prescription {
  attribute Double honoraire;
  attribute Struct T_Tension{
                      Short max,
                      Short min} tension;
  attribute Double poids;
  attribute Double taille;
  relationship Praticien prescripteur
      inverse Praticien::consultations;
  attribute Set<Struct> T_medicament {
            String code,
            Boolean generique,
            String categorie_molecule,
            String type_molecule,
            Short quantite,
```

```
          Double tarif}> medicament;
 Double montant_euro();
 Double montant_prescrit();
```

## 8.4 Magasin de données

A partir de l'entrepôt, il est alors possible de construire des schémas multidimensionnels ; nous avons étudié ce processus de transformation dans [BRET99] [BRET00].

L'administrateur peut construire un magasin pour supporter par exemple des analyses relatives aux honoraires des praticiens ainsi qu'au nombre de médicaments prescrits et au nombre de médicaments génériques prescrits.

La Figure 12 décrit le schéma étoile supportant les analyses relatives aux taux de médicaments génériques prescrits en fonction de la répartition géographique ou par catégorie, puis spécialité de praticiens.

```
String affiche_tension();
Double cout_secu(); }
```

- **<u>Magasin de données</u> :**

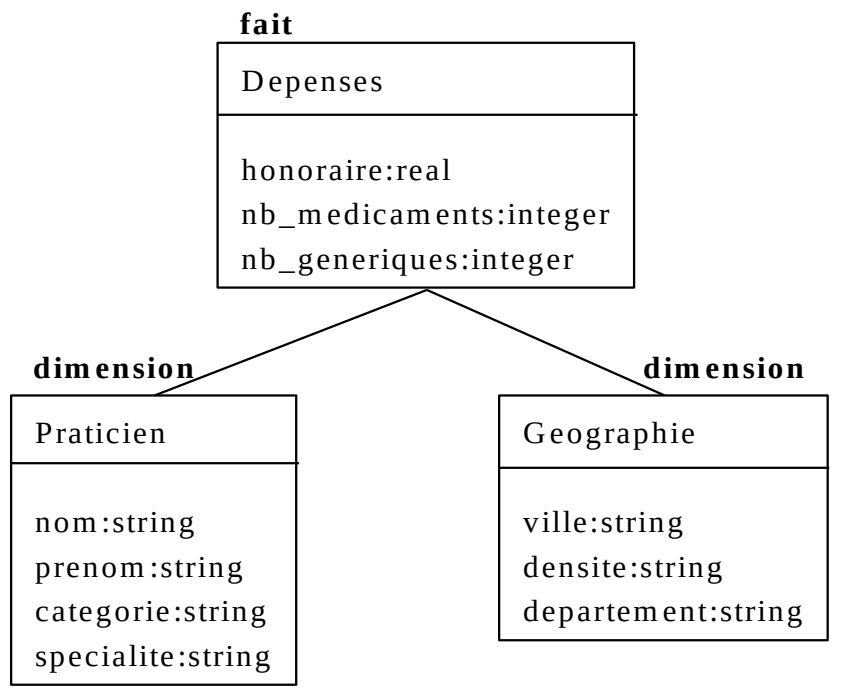

**Figure 12 :** Exemple d'un schéma en étoile.